# Pump Pulse Bandwidth-Activated Nonlinear Phononic Coupling in CdWO$_4$

Megan F. Biggs[1]†, Brittany E. Knighton[1]†, Aldair Alejandro[1]†, Lauren M. Davis[1]†, Claire Rader[1], Jeremy A. Johnson[1]*

[1]Department of Chemistry and Biochemistry, Brigham Young University; Provo, 84602 USA

*Corresponding author. Email: jjohnson@chem.byu.edu

† These authors contributed equally to this work

**Abstract:** To control structure-function relationships in solids with light, we must harness the shape of the potential energy surface, as expressed in anharmonic coupling coefficients. We use two-dimensional terahertz (THz) spectroscopy to identify trilinear coupling between sets of vibrational modes in CdWO$_4$. It is generally understood that efficient trilinear coupling occurs when the frequencies of two coupled modes add or subtract to the frequency of the third mode. Interestingly, we observe that this condition is not necessary: the THz driving-pulse itself can activate the coupling by contributing broad frequency content to the initial motion of the excited modes. Understanding that the bandwidth of the driving force can activate energy-flow pathways has broad implications for coherent control of collective modes using intense THz light pulses.

**One-Sentence Summary:** A broadband terahertz pump pulse can "turn on" energy-flow pathways, affecting the control of structure-function relationships with light.

**Main Text:** The ability to use light to alter material properties or induce new functionality "on demand" (*1, 2*) is closely connected to the nascent field of nonlinear phononics (*3-12*). The central tenet of chemistry and materials science is, after all, that structure determines the function and properties of materials. Nonlinear phononics provides the path to harness the fundamental forces that bind materials together. We use ultrafast laser pulses to launch atoms on useful trajectories; trajectories that can alter the structure and associated material properties. For example, recently, terahertz (THz) and mid-infrared light have been used to induce ferroelectricity in $SrTiO_3$ (*4,11*), enhance superconductivity (*13*), and transiently reverse the ferroelectric polarization of $LiNbO_3$ (*5*).

The fundamental forces that govern ultrafast structural manipulation are directly connected to a material's potential energy surface (PES). In crystalline materials, the atomic motions for each unit cell can be decomposed into normal vibrational modes that constitute zone-center phonon modes. As the atoms are displaced from their equilibrium positions along normal-mode coordinates, the potential energy initially rises quadratically according to the harmonic approximation. As atoms move larger distances from equilibrium, anharmonicities start to exhibit that can be described by higher order (than quadratic) terms in the PES. Moreover, the addition of an external force like an electric field or incident light pulse can further affect the PES. Equation 1 describes the PES for a material with *n* vibrational modes ($Q_i$) in the absence of external fields.

$$V(\mathbf{Q}) = \sum_{i=1}^{n} \frac{1}{2}\omega_i^2 Q_i^2 + \sum_i \sum_j \sum_k \phi_{ijk} Q_i Q_j Q_k + \cdots \qquad \text{Eq. 1}$$

where $\omega_i = 2\pi\nu_i$ is the angular frequency of each mode, and $\nu_i$ is the real frequency. $\phi_{ijk}$ are third-order force constants that determine the anharmonic shape of the PES and govern anharmonic coupling between modes. Before introducing the relevant equations of motion that are derived from Eq. 1, we will introduce some properties of $CdWO_4$, which is the material of focus of this work.

$CdWO_4$ is a centrosymmetric crystal with four $B_g$ Raman-active and three IR(infrared)-active ($A_u$ and $B_u$) phonon modes within the frequency range that we can excite and probe (0.5 to 4.5 THz) (Table 1) (*8, 14*).

**Table 1. $CdWO_4$ Vibrational Modes**

| Band Index | $\nu_i$ (THz) | Symmetry | Raman/IR |
|---|---|---|---|
| 1-3 | - | Acoustic | Acoustic |
| 4 | 2.33 | $B_g$ | R |
| 5 | 2.90 | $A_g$ | R |
| 6 | 2.94 | $B_u$ | IR |
| 7 | 3.56 | $B_g$ | R |
| 8 | 3.65 | $A_u$ | IR |
| 9 | 4.04 | $B_g$ | R |
| 10 | 4.47 | $B_u$ | IR |
| 11 | 4.48 | $B_g$ | R |

**Coupled modes and nonlinear excitation**

We can use Eq. 1 to derive equations of motion for Raman and IR-active modes in CdWO4.

$$\ddot{Q}_j + 2\Gamma_j\dot{Q}_j + \omega_j^2 Q_j = Z_{j\alpha}^* E_\alpha - \sum_i \sum_k \phi_{ijk} Q_i Q_k + \cdots \quad \text{Eq. 2}$$

$$\ddot{Q}_i + 2\Gamma_i\dot{Q}_i + \omega_i^2 Q_i = -\epsilon_0 R_{i\alpha\beta} E_\alpha E_\beta - \sum_j \sum_k \phi_{ijk} Q_j Q_k + \cdots \quad \text{Eq. 3}$$

Eq. 2 shows the excitation of an IR-active mode ($Q_j$) through direct driving by the pump field ($Z_{j\alpha}^* E_\alpha$) or third-order coupling ($\sum_i \sum_k \phi_{ijk} Q_i Q_k$), while Eq. 3 shows the excitation of a Raman-active mode ($Q_i$) through Raman excitation ($-\epsilon_0 R_{i\alpha\beta} E_\alpha E_\beta$) or third-order coupling. In these equations, $\Gamma_i$ represents the damping rate, $Z_{j\alpha}^*$ is the mode effective charge (which is non-zero for IR-active modes), $\epsilon_0$ is the permittivity of free space, $E_\alpha$ and $E_\beta$ are THz electric fields ($\alpha$ and $\beta$ are THz polarization directions), and $R_{i\alpha\beta}$ is the Raman tensor element. According to the equations of motion, different pathways can lead to phonon excitation (*8,15*). If a mode is IR-active, that mode can resonantly be excited directly with THz electric field. If a mode is Raman-active, then it can be excited through a two-field interaction Raman pathway (*16*). Additionally, if other modes ($Q_j$ and $Q_k$) are excited to large amplitudes, then a third mode ($Q_i$) can be excited via anharmonic coupling (*15,17*). Symmetry considerations indicate that third-order coupling in CdWO4 involving a B$_g$ Raman-active mode, must also include an A$_u$ IR-active mode and a B$_u$ IR-active mode (further explained in the Supplemental Material S1).

Excitation of a particular mode $Q_i$ will be most efficient when the driving force terms on the right side of Eq. 2 or 3 contain large amplitudes at the resonant frequency of mode $Q_i$. For resonant excitation (Eq. 2), this means the THz driving pulse has significant spectral amplitude at the resonant frequency. The driving force terms included in Eq. 3 are third-order processes like anharmonic coupling and electronic Raman scattering (ERS) that include products of electric fields ($E_\alpha E_\beta$) or vibrational modes ($Q_j Q_k$). These products mean that the resulting driving force from third-order nonlinear pathways will have the largest effect when sums or differences between frequency components of fields and mode motions match the frequency of mode $Q_i$ (*15,17*). For example, in trilinear coupling terms, if two phonon modes $Q_j$ and $Q_k$ are resonantly excited, they will oscillate at their resonant frequencies. If we model that resonant oscillatory motion, then the combined motion that is the anharmonic coupling driving force term in Eq. 3 has the form

$$\phi_{ijk} Q_j Q_k = \phi_{ijk} \cos(\omega_j t)\cos(\omega_k t) = \phi_{ijk} \frac{1}{2}[\cos((\omega_j - \omega_k)t) + \cos((\omega_j + \omega_k)t)]. \quad \text{Eq. 4}$$

Therefore, anharmonic coupling is most efficient when the resonant frequencies of the two directly excited phonons add up or subtract down to the frequency of the coupled mode $Q_i$. An equivalent criterion is present in ERS, where the frequencies of the interacting fields $E_\alpha, E_\beta$ must add or subtract to the frequency of the vibrational mode that is nonlinearly excited (*15*).

Using two-dimensional (2D) THz spectroscopy (*8,18-22*) (see also Materials and Methods in Supplemental Material), we identify clear signals that indicate two sets of third-order anharmonic couplings in CdWO4 lead to excitation of the B$_g$ mode 4 at 2.33 THz: couplings between modes 4, 6, and 8 ($\phi_{4,6,8}$) and between modes 4, 8, and 10 ($\phi_{4,8,10}$). Interestingly, the resonant frequencies of the IR-active modes involved in these coupling processes do not add or subtract to the 2.33 THz frequency of the excited mode 4. We show that the bandwidth of the pump pulses activates these anharmonic couplings so that Raman-active modes can be excited.

**2D spectroscopy identifies specific anharmonic couplings**

As depicted in Fig. 1A and 1B, we experimentally isolate nonlinear signals in a 2D THz experiment using a differential chopping scheme. Two THz pump pulses with a variable relative delay ($\tau$) and an 800-nm Raman probe (delay $t$) are focused onto the sample (Fig. 1A). Choppers are placed in both THz beam paths to produce a sequence of pump pulses with every four laser shots, allowing us to perform simple subtraction to isolate third-order nonlinear signals (Fig. 1B). Fig. 1C shows the nonlinear signal from $CdWO_4$ for a range of delays between the THz pump pulses. Perpendicularly polarized THz pulses minimizes signal arising from ERS and enables clear identification of anharmonic coupling (Supplemental Material S6). Once the nonlinear signal is isolated in time, the Fourier transform is taken along both time axes to produce a 2D frequency correlation plot (Fig. 2).

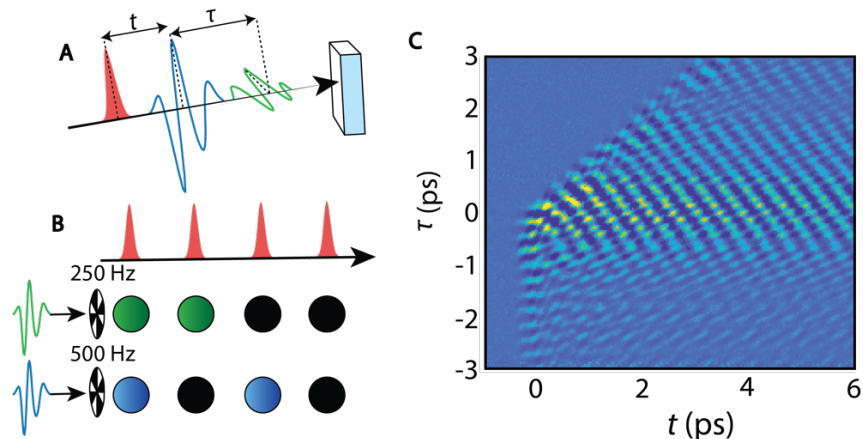

**Fig. 1. Overview of 2D THz spectroscopy.** (**A**) The two perpendicularly polarized THz pump pulses (blue, green) with delay between them ($\tau$) and the 800 nm probe pulse (red) with delay $t$. (**B**) Dual chopping scheme. The 500 Hz chopper blocks every other pump pulse. The 250 Hz chopper blocks two pulses and lets two through, giving a pattern of four pump combinations: both THz pulses, vertical THz alone, horizontal THz alone, and no pump pulses. (**C**) Isolated 2D nonlinear signal as a function of probe delay $t$ (x-axis) and pump delay $\tau$ (y-axis).

In Fig. 2, features appear at frequencies along the probe frequency axis that correspond to Raman-active modes that are nonlinearly excited and probed with our 800-nm probe pulse. For example, we see a vertical stripe of features that appear at 2.33 THz along the probe axis that correspond to $B_g$ Raman-active mode 4. A long, thin stripe of signal along this vertical line indicates that THz Raman excitation of this mode is occurring (Supplemental Material S6). Additional bright features along this vertical line, identified with white and gray shapes, arise from third-order anharmonic coupling between the 2.33 THz $B_g$ mode and resonantly driven IR-active modes. Symmetry considerations require that this trilinear third-order coupling includes the excited $B_g$ Raman-active mode as well as one $A_u$ IR-active mode and one $B_u$ IR-active mode (Supplemental Material S1).

On the pump axis, features may appear at the resonant frequencies of the IR-active modes (circles) as well as at the sum- and difference-frequencies between the Raman-active mode and the IR-active modes, diamonds and squares respectively (Supplemental Material S7). Diamonds, circles, and squares in Fig. 2 highlight the nonlinear signals arising from trilinear coupling between modes 4, 6, and 8 (gray), and between modes 4, 8, and 10 (white).

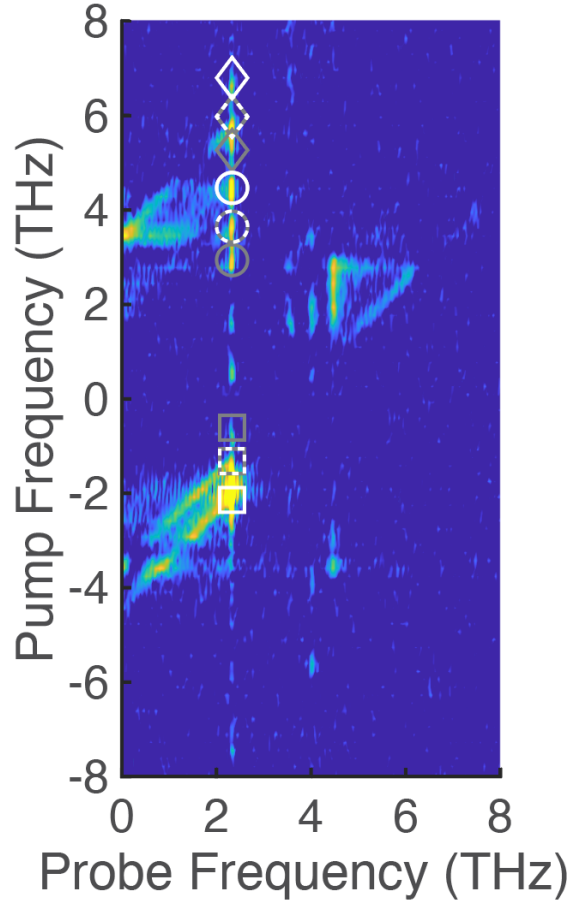

**Fig. 2. 2D Frequency correlation spectrum of the nonlinear signal in CdWO$_4$ with coupling signals indicated.** Coupling signals are marked showing features at IR-active mode resonant frequencies (circles), sum-frequencies (diamonds), and difference-frequencies (squares) arising from the coupling between modes 4,6,8 (gray) and 4,8,10 (white). Signals that arise from both couplings are indicated with dashed gray and white lines.

If we examine the coupling between modes 4, 8, and 10, we see features that appear at the frequency of the Raman-active mode 4 (at 2.33 THz) on the probe axis and at the frequencies of the two IR-active modes on the pump axis (3.65 THz and 4.47 THz) (marked with white circles on Fig. 2). We also see the sum-frequency signals shown in Fig. 2 as white diamonds at 5.98 THz (2.33 THz + 3.65 THz) and 6.80 THz (2.33 THz + 4.47 THz). Coupling between modes 4, 8, and 10 also results in difference-frequency signals at -1.32 THz (2.33 THz - 3.65 THz) and -2.14 THz (2.33 THz - 4.47 THz) (marked with white squares). A similar pattern of signals is indicated with gray markers on Fig. 2 for coupling between modes 4, 6, and 8.

**Experimental couplings are compared to first-principles calculations**

To quantify the amplitudes of these couplings, we extract coupling constants $\phi_{4,6,8}$ and $\phi_{4,8,10}$ by modelling the vibrational motion to produce frequency correlation plots. To model the vibrational motion, we solve the coupled equations of motion (Eq. 2 and 3) for all vibrational modes, using the electric field from measured experimental THz traces as $E_\alpha$ and $E_\beta$. We can then model the motion for the three nonzero THz pump pulse combinations shown in Fig. 1B. This allows us to model the 2D nonlinear time signal corresponding to the data in Fig. 1C. We then Fourier transform along both axes to produce a modeled 2D frequency correlation plot. To extract $\phi_{468}$, $\phi_{4610}$, and $R_{xz4}$ from our experimental data, we perform a manual chi-squared fitting which compares the modeled frequency correlation plot to the experimental frequency correlation plot. We change all the parameter values ($\phi_{4,6,8}$, $\phi_{4,6,10}$, and $R_{xz4}$) until the root-mean square-error (RMSE) of the difference between the modeled and experimental spectra was minimized (see the Supplemental Material S5 for more details). We also use first-principles calculations to determine $\phi_{4,6,8}$ and $\phi_{4,8,10}$ (Supplemental Material S8).

Due to the fact that the experimental signal is proportional to the displacement of the Raman-active modes, and actual atomic displacements are not measured, the chi-squared fit can only extract relative values for coupling constants and Raman tensor elements. Therefore, to directly compare the coupling constants extracted from experimental data to the first-principles values, we compare the ratio $\phi_{4,8,10}/\phi_{4,6,8}$. The value calculated with first-principles calculations of $\phi_{4,8,10}/\phi_{4,6,8} = -0.67$ compares favorably (sign and magnitude) to the experimentally determined ratio of $-1.23$.

**THz bandwidth 'turns on' inefficient couplings**

The 2D THz data clearly shows evidence for anharmonic excitation of Raman-active modes. However, for the couplings that we have identified in CdWO$_4$, the frequencies of the IR-active modes involved do not add or subtract to a value close to the frequency of the excited Raman-active mode (Table 2), as we would expect from Eq. 4 for efficient nonlinear excitation. To understand how anharmonic coupling still occurs with these inefficient resonant frequency combinations, let us carefully examine the coupling between modes 4, 8, and 10.

**Table 2. Frequency Combinations of Important Trilinear Couplings (in THz)**

| Coupled Modes | $v_j$ (mode IR$_1$) | $v_k$ (mode IR$_2$) | Sum | Diff. | Diff. needed | $v_i$ (mode R) |
|---|---|---|---|---|---|---|
| 4, 6, 8 | 2.94 (mode 6) | 3.65 (mode 8) | 6.59 | 0.71 | 1.32 | 2.33 (mode 4) |
| 4, 8, 10 | 3.65 (mode 8) | 4.47 (mode 10) | 8.12 | 0.82 | 2.14 | 2.33 (mode 4) |

In 2D THz measurements, we change the arrival timing between a pair of THz pulses, one "variable delay" horizontally polarized pulse that arrives at the sample at a delay ($\tau$) relative to a "stationary" vertically polarized pulse. Fig. 3 illustrates an example of a specific pump delay ($\tau$) when the variable delay pulse, polarized to excite A$_u$ mode 8 (3.65 THz), arrives a few picoseconds before the stationary pulse, polarized to excite B$_u$ mode 10 (4.47 THz). The top panels (A and C) of Fig. 3 show the modeled atomic motion of mode 8 and mode 10 when resonantly excited with appropriate delay. The atomic motion contains more frequencies when the THz pump pulse is present compared to later times: when an oscillator is in the process of being driven, frequency components from the driving force will be imparted to the transient motion of the oscillator, even if the force is not resonant with the mode. After the driving force has ended, the mode will oscillate at its resonant frequency.

As an example, the Fourier transform shown in Fig. 3B shows the narrowband resonant frequency in the late-time oscillations highlighted in red in Fig. 3A. Fig. 3D shows the broadband frequency content of the motion of mode 10 while it is being driven by the second THz pulse, indicated in purple in Fig. 3C. Therefore, when the second THz pulse arrives a few picoseconds after the first, the atomic motion in mode 8 only contains the resonant frequency of the mode (3.65 THz). On the other hand, when mode 10 is being driven by the second THz pulse, the early motion of the mode contains all of the frequencies of the THz pump pulse. The broad frequency range in the early motion makes it possible to subtract from 3.65 THz down to the frequency of the Raman-active mode at 2.33 THz; the early motion of mode 10 must simply contain a frequency component at 1.32 THz (3.65 THz – 2.33 THz) to activate the anharmonic coupling. As shown in Fig. 3D with the vertical green dotted line, the early atomic motion of mode 10 with its broad spectrum includes the required component at 1.32 THz, making this coupling to mode 4 possible. Although we only show an example for one THz pulse delay $\tau$, the THz bandwidth activation occurs throughout the 2D measurement for both the $\phi_{4,8,10}$ and $\phi_{4,6,8}$ couplings.

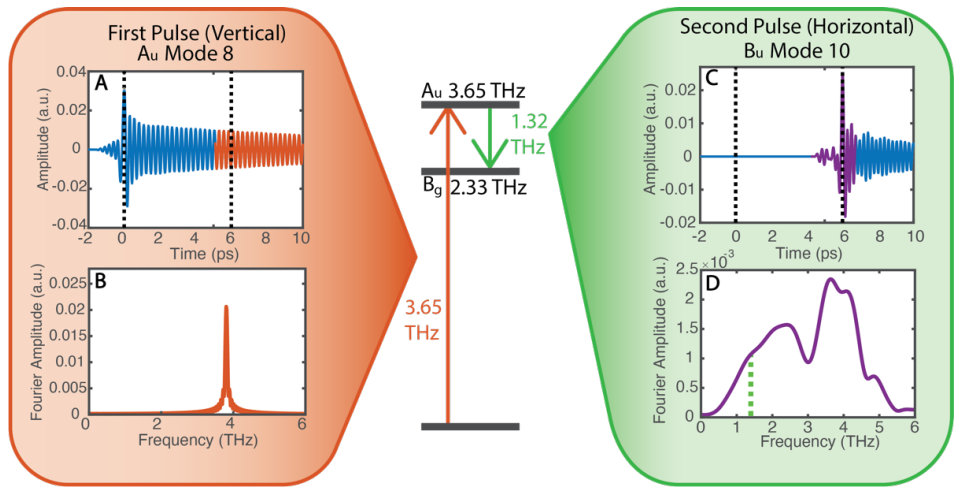

**Fig. 3. The THz pump pulse can contribute frequency content to vibrational motion, enabling trilinear coupling.** The energy diagram in the center of the figure indicates the phonon difference-frequency combination required for $A_u$ mode 8 (orange arrow) and $B_u$ mode 10 (green arrow) to trilinearly couple to and excite $B_g$ mode 4. The orange box on the left shows modeled motion of $A_u$ mode 8 when it has been excited 6 ps before $B_u$ mode 10, which is modeled in the green box on the right. Vertical dotted black lines in **A** and **C** indicate arrival of THz pump pulses. (**B**) The Fourier transform of the late atomic motion of mode 8 (red-orange motion in **A**), showing a peak at the resonant frequency of the mode. (**D**) Fourier transform of the early motion of mode 10 (purple in **C**) indicating that the early atomic mode motion includes all the frequencies of the THz pump pulse when the pump pulse is present. The green dotted line in **D** indicates the frequency required to subtract down from mode 8 to the frequency of mode 4.

In conclusion, we have used 2D THz spectroscopy to identify third-order anharmonic couplings in CdWO$_4$. By modeling the 2D frequency correlation spectrum using coupled equations of motion, we quantitatively determined ratios of coupling constants, which compare favorably with those calculated via 1$^{st}$ principles calculations. Perhaps the most important result is that we observe anharmonic coupling pathways can be activated by the bandwidth of the broadband THz driving pulses.

The existence of nonlinear excitation pathways that are activated while modes are being driven means that simple models might not reveal all the complicated energy pathways within a material. A simple model involving only sum- and difference-frequencies would still function for narrowband, resonant vibrational motion, but when broad-band driving pulses are present a more complex model must be used to account for all possible energy-flow pathways.

**Acknowledgments:**

**Funding:** The authors acknowledge funding from the Simmons Research Fund as well as the Department of Chemistry and Biochemistry and the College of Physical and Mathematical Sciences at Brigham Young University.

## Supplementary Materials

**Materials and Methods**

Methods

Two-dimensional (2D) time-domain terahertz (THz) spectroscopy was used to examine nonlinear excitations in a (010) 500-μm-thick CdWO$_4$ crystal. In these 2D measurements, optical-parametric-amplifier-(OPA)-generated signal (1450 nm) and idler (1790 nm) 100 fs pulses were directed into separate DAST THz generation crystals to produce two pump pulses with a frequency content between 0.5 THz and 5 THz. The THz pulses were directed through a three-parabolic-mirror focusing setup onto the sample (see Figure S1 and S3). Each THz beam is approximately 250 μm in diameter when overlapped at the focus.

An 800-nm, 100-fs probe generated from the Ti:Sapphire laser was focused onto the sample and spatiotemporally overlapped with the focus of the two THz pump beams. The magnitude of the excited vibrational motion was measured using heterodyne detection and the same sensitive detection scheme described in Ref. (*8, 23, 24*). The timing of the "variable delay" THz pulse (generated from the OPA idler output) with respect to the "stationary" THz pulse (generated from the OPA signal output) was varied over the course of the experiment using a delay stage with a 0.005 mm step size.

The azimuthal angle of CdWO$_4$ was oriented at 0 degrees, such that the variable delayed THz pulse was horizontally polarized to excite A$_u$ symmetry modes and the stationary pulse was vertically polarized to excite B$_u$ symmetry modes. Both generated pulses were overlapped in space using a wire-grid polarizer. The polarizer was set to transmit horizontally polarized THz pulses and reflect vertically polarized THz pulses.

The repetition rate of the laser throughout these experiments was 1 kHz. A dual chopping scheme simplified data processing by allowing the isolation of different specific signals of interest (see Fig. 1B of the main manuscript). A chopper with a 500 Hz frequency was placed in the OPA signal output path (stationary THz pump pulse) and a chopper with a 250 Hz frequency was placed in the OPA idler output path (variably delayed THz pump pulse). This chopping scheme allowed us to record signals with four different THz pump configurations: both pulses blocked, both pulses unblocked, only variably delayed THz pulse unblocked, and only stationary THz pulse unblocked. Through a simple subtraction shown in Eq. S1, the nonlinear signal could then be isolated.

$$S_{nonlinear} = S_{both} - S_{stationary\ pump} - S_{variable\ pump} \qquad \text{Eq. S1}$$

**Supplementary Text**

S1 Group Theory and Excitation Pathways

Within our THz bandwidth (0.5 - 5 THz), there are three IR-active phonons (Table S1) and five Raman-active phonons (Table S2) in CdWO$_4$. A THz photon can directly excite IR-active phonons if the photon has the appropriate energy and polarization. For the y-cut CdWO$_4$ crystal in our experiments, the 800-nm probe and THz pump travel parallel to the y-crystallographic direction, with relevant polarizations in the x-z plane. The three IR-active modes are of either A$_u$ or B$_u$ symmetry. According to the 6$^{th}$ column of the *C$_{2h}$* character table (Table S3), modes of A$_u$ symmetry can be directly excited by pump light polarized in the z direction while modes of $B_u$ symmetry can be directly excited by pump light polarized in the x or y direction.

| Band Index | Frequency | Symmetry |
|---|---|---|
| 6 | 2.94 THz | $B_u$ |
| 8 | 3.65 THz | $A_u$ |
| 10 | 4.47 THz | $B_u$ |

**Table S1 IR-Active Phonon Modes (from Ref. (*14*))**

| Band Index | Frequency (Experimental) | Symmetry |
|---|---|---|
| 4 | 2.33 THz | $B_g$ |
| 5 | 2.90 THz | $A_g$ |
| 7 | 3.56 THz | $B_g$ |
| 9 | 4.04 THz | $B_g$ |
| 11 | 4.48 THz | $B_g$ |

**Table S2 Raman-Active Phonon Modes**

| $C_{2h}$ | E | $C_2$ | i | $\sigma$ | | |
|---|---|---|---|---|---|---|
| $A_g$ | 1 | 1 | 1 | 1 | $R_z$ | $x^2, y^2, z^2, xy$ |
| $B_g$ | 1 | -1 | 1 | -1 | $R_x, R_y$ | $xz, yz$ |
| $A_u$ | 1 | 1 | -1 | -1 | z | |
| $B_u$ | 1 | -1 | -1 | 1 | x, y | |

**Table S3 $C_{2h}$ Character Table**

| $\otimes$ | $A_g$ | $A_u$ | $B_g$ | $B_u$ |
|---|---|---|---|---|
| $A_g$ | $A_g$ | $A_u$ | $B_g$ | $B_u$ |
| $A_u$ | | $A_g$ | $B_u$ | $B_g$ |
| $B_g$ | | | $A_g$ | $A_u$ |
| $B_u$ | | | | $A_g$ |

**Table S4 Symmetry Multiplication**

Calculating the direct product using Table S4 determines if the excitation is symmetry-allowed. For example, in the case of direct excitation of an $A_u$ IR-active phonon mode, the oscillator begins in the totally symmetric ground state as $A_g$, interacts with the pump electric field with a z polarization ($\hat{\mu}_z$) and ends excited in the $A_u$ state.

$$\langle \psi_f | \hat{\mu} | \psi_i \rangle \propto \langle A_u | \hat{\mu}_z | A_g \rangle \propto \langle A_u | A_u | A_g \rangle \supset A_g \tag{S2}$$

The transition is allowed if the product contains $A_g$, as is the case in Eq. S2. Table S5 summarizes which interactions are symmetry-allowed and symmetry-forbidden in CdWO$_4$ considering single-photon, two-photon, and trilinear phonon coupling mechanisms as driving forces for IR and Raman-active modes. This information shows us both electronic Raman scattering (ERS) and trilinear phonon coupling can be symmetry-allowed and must be considered as possible excitation sources. To distinguish between these two mechanisms, we model the 2D spectrum based on the unique equations of motion from each process.

| Mode Type | Excitation Type | Direct Products |
|---|---|---|
| IR-Active mode | Single Photon (Resonant Excitation) | $\langle A_u\|\hat{u}_z\|A_g\rangle \propto \langle A_u\|A_u\|A_g\rangle \supset A_g$ <br> Allowed <br> $\langle B_u\|\hat{u}_x\|A_g\rangle \propto \langle B_u\|B_u\|A_g\rangle \supset A_g$ <br> Allowed |
| Raman-Active mode | Single Photon (Resonant Excitation) | $\langle A_g\|\hat{u}_z\|A_g\rangle \propto \langle A_g\|A_u\|A_g\rangle \not\supset A_g$ <br> $\langle A_g\|\hat{u}_x\|A_g\rangle \propto \langle A_g\|B_u\|A_g\rangle \not\supset A_g$ <br> Forbidden: Product does not contain $A_g$ <br> $\langle B_g\|\hat{u}_z\|A_g\rangle \propto \langle B_g\|A_u\|A_g\rangle \not\supset A_g$ <br> $\langle B_g\|\hat{u}_x\|A_g\rangle \propto \langle B_g\|B_u\|A_g\rangle \not\supset A_g$ <br> Forbidden: Product does not contain $A_g$ |
| Raman-Active mode | Single Photon (Hot Band Excitation) | $\langle B_g\|\hat{u}_z\|B_u\rangle \propto \langle B_g\|A_u\|B_u\rangle \supset A_g$ <br> Allowed |
| Raman-Active mode | Two-Photon (Electronic Raman Scattering) | $\langle B_g\|\hat{u}_z\hat{u}_x\|A_g\rangle \propto \langle B_g\|A_uB_u\|A_g\rangle \supset A_g$ <br> Allowed <br> $\langle B_g\|\hat{u}_z\hat{u}_z\|A_g\rangle \propto \langle B_g\|A_uA_u\|A_g\rangle \not\supset A_g$ <br> Forbidden: Product does not contain $A_g$ <br> $\langle B_g\|\hat{u}_x\hat{u}_x\|A_g\rangle \propto \langle B_g\|B_uB_u\|A_g\rangle \not\supset A_g$ <br> Forbidden: Product does not contain $A_g$ |
| Raman-Active mode | Trilinear phonon coupling | $\langle B_g\|A_uB_u\rangle \supset A_g$ <br> Allowed <br> $\langle B_g\|B_uB_u\rangle \not\supset A_g$ <br> Forbidden: Product does not contain $A_g$ <br> $\langle B_g\|A_uA_u\rangle \not\supset A_g$ <br> Forbidden: Product does not contain $A_g$ |

**Table S5 Direct Product Summary**

S2 Equations of Motion

$$\ddot{Q}_{IR_8} + 2\Gamma_{IR_8}\dot{Q}_{IR_8} + \omega_{IR_8}^2 Q_{IR_8} = Z_{8\alpha}^* E_\alpha - \sum_{i=4,7,9,11} \phi_{i,6,8} Q_{IR_6} Q_{R_i} - \sum_{i=4,7,9,11} \phi_{i,8,10} Q_{IR_{10}} Q_{R_i}$$

$$\ddot{Q}_{IR_6} + 2\Gamma_{IR_6}\dot{Q}_{IR_6} + \omega_{IR_6}^2 Q_{IR_6} = Z_{6\beta}^* E_\beta - \sum_{i=4,7,9,11} \phi_{i,6,8} Q_{IR_8} Q_{R_i} \quad (S2)$$

$$\ddot{Q}_{IR_{10}} + 2\Gamma_{IR_{10}}\dot{Q}_{IR_{10}} + \omega_{IR_{10}}^2 Q_{IR_{10}} = Z_{10\beta}^* E_\beta - \sum_{i=4,7,9,11} \phi_{i,8,10} Q_{IR_8} Q_{R_i}$$

The equations of motion quantify how atomic forces interact to describe phonon motion and are derived from the potential energy surface equation. In the equations of motion above, the left side of the equation describes the forces governing a damped harmonic oscillator and the right side describes the driving force. Equations S2 describe excitation of the three IR-active modes. We expect that the back-coupling terms (the summation terms) do not make any noticeable contribution to the signal because of the small amplitudes of the Raman-active modes, and we omit them in our final modeling process. $Q_j$ is the phonon mode coordinate, $\Gamma_j$ is the phonon damping rate, $\omega_j$ is the mode frequency (angular frequency), $Z_{j\alpha}^*$ is the mode effective charge, which determines how strongly the electric field ($E_\alpha$) interacts with $Q_j$, and $\phi_{i,j,k}$ is the coupling constant, which relates how strongly two other phonons couple to the motion of $Q_j$. $E_\alpha$ and $E_\beta$ are a vertically polarized and horizontally polarized electric fields respectively. $E_\alpha$ is oriented to excite $A_u$ modes and $E_\beta$ is oriented to excite $B_u$ modes.

The Raman modes have two symmetry allowed excitation mechanisms, and as a result both ERS ($R_{i\alpha\beta}E_\alpha E_\beta$) and trilinear phonon coupling ($\phi_{i,j,k}Q_{IR_j}Q_{IR_k}$) are included as the driving forces in the equations of motion (Equations S3).

$$\ddot{Q}_{R_4} + 2\Gamma_{R_4}\dot{Q}_{R_4} + \omega_{R_4}^2 Q_{R_4} = \epsilon_0 R_{4\alpha\beta} E_\alpha E_\beta - \phi_{4,6,8} Q_{IR_6} Q_{IR_8} - \phi_{4,8,10} Q_{IR_8} Q_{IR_{10}}$$

$$\ddot{Q}_{R_7} + 2\Gamma_{R_7}\dot{Q}_{R_7} + \omega_{R_7}^2 Q_{R_7} = \epsilon_0 R_{7\alpha\beta} E_\alpha E_\beta - \phi_{7,6,8} Q_{IR_6} Q_{IR_8} - \phi_{7,8,10} Q_{IR_8} Q_{IR_{10}}$$

$$\ddot{Q}_{R_9} + 2\Gamma_{R_9}\dot{Q}_{R_9} + \omega_{R_9}^2 Q_{R_9} = \epsilon_0 R_{9\alpha\beta} E_\alpha E_\beta - \phi_{9,6,8} Q_{IR_6} Q_{IR_8} - \phi_{9,8,10} Q_{IR_8} Q_{IR_{10}} \quad (S3)$$

$$\ddot{Q}_{R_{11}} + 2\Gamma_{R_{11}}\dot{Q}_{R_{11}} + \omega_{R_{11}}^2 Q_{R_{11}} = \epsilon_0 R_{11\alpha\beta} E_\alpha E_\beta - \phi_{11,6,8} Q_{IR_6} Q_{IR_8} - \phi_{11,8,10} Q_{IR_8} Q_{IR_{10}}$$

Unlike the resonant excitation process, ERS requires two electric field interactions and by symmetry rules, these two fields must have perpendicular polarizations. We can achieve this configuration by having THz linearly polarized at some diagonal angle between the x and z axis, or one THz pulse along the x axis and the other THz pulse along the z axis. For the case of symmetry-allowed trilinear-phonon coupling, two driven IR-active modes are coupled to a single Raman-active mode, and the resulting force is dependent on the phonon mode coordinates ($Q_{IR_j}, Q_{IR_k}$) and the coupling coefficient ($\phi_{i,a,b}$). For trilinear coupling to be symmetry allowed, the coupling must involve one $A_u$ IR-active mode, one $B_u$ IR-active mode, and one $B_g$ Raman-active mode.

S3 Experimental Considerations
*__THz pump pulses with Parallel Polarization__*

Fig. S1 shows the experimental setup when the two THz pump pulses are both vertically polarized. The "stationary" pulse is generated by the signal output pulse from the optical parametric amplifier (OPA), which is vertically polarized. The idler output from the OPA is used to create the "moving" pulse. The idler exits the OPA beam splitter horizontally polarized. A HWP is placed in the idler path at 45º to rotate the idler polarization to vertical before arriving at the DAST THz generation crystal. A d-mirror reflects the stationary THz pulse into the parabolic mirror scheme while the moving pulse is directed above the d-mirror into the parabolic mirrors. They are adjusted to be as parallel as possible when they reflect off the final parabolic mirror, allowing them to overlap spatially at the focus. The $CdWO_4$ sample is rotated 45º from the optical axis.

This parallel polarization configuration allows the electric field of the two THz pulses to excite the Raman-active phonons by ERS at all THz delays because each individual pulse contains multiple photons with the required orthogonal symmetries to excite a mode. Additionally, each THz pulse can be divided into polarization components along both IR-active modes involved in trilinear phonon coupling ($B_u$ and $A_u$). Fig. S2 shows the 2D plots resulting from the parallel THz measurement. In Fig. S2A, $CdWO_4$ was pumped by both THz pulses. The vertical lines are the sample response to the stationary THz pulse and the diagonal lines are from the sample response to the variable-delay THz pulse. When the two pulses are overlapped in the sample, the sample response is more complex than just the sum of the stationary and moving signals. The isolated nonlinear signal in Fig. S2B is converted into the frequency domain and results in the feature-rich nonlinear 2D frequency correlation spectrum in Fig. S2C.

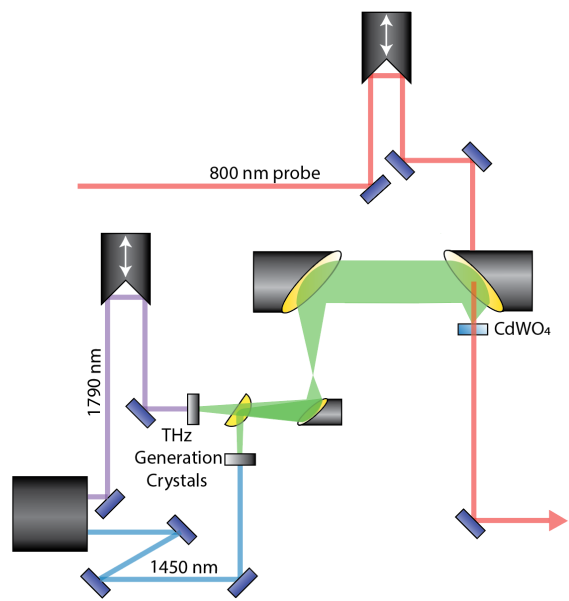

**Fig. S1. Experimental setup to perform 2D THz measurements with parallel pump polarizations.** Experimental setup showing the signal (blue, 1450 nm) and idler (purple, 1790 nm) OPA outputs traveling through DAST crystals to generate THz (green). The signal THz pulses reflect off a d-mirror while the idler travels over the d-mirror. The two THz pulses travel parallel through the parabolic mirror scheme to focus to the same spot on the $CdWO_4$ sample. The 800 nm probe pulse (red) is focused to the same spot on the sample and measured through heterodyne detection.

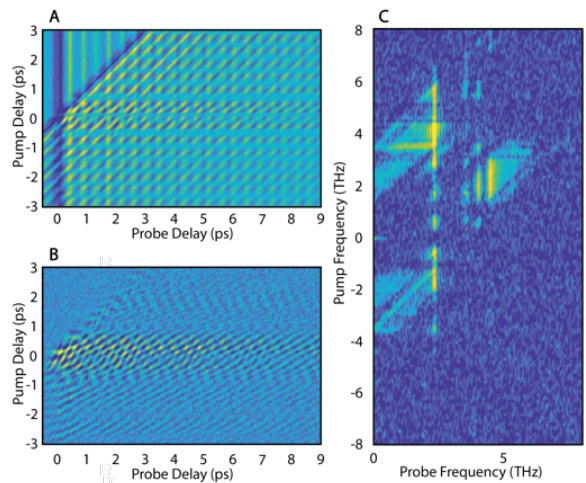

**Fig. S2. Time and frequency signals from 2D THz spectroscopy with parallel pump polarizations.** (**A**) A 2D plot of the time domain signal where the vertical lines are from the stationary THz pulse and the diagonal lines are from the moving THz pulse. (**B**) The nonlinear time-domain signal isolated after chopping and (**C**) the resulting 2D frequency correlation spectrum.

*THz pump pulses with Perpendicular Polarization*

Fig. S3 shows the experimental setup when the two THz pump pulses have perpendicular polarizations. The signal OPA output generates vertically polarized THz while the idler output generates horizontally polarized THz when in contact with the DAST generation crystal. A polarizer oriented to transmit horizontally polarized light reflects the vertically polarized signal THz pulse into the parabolic mirror scheme while the horizontally polarized idler THz pulse travels through the polarizer into the parabolic mirrors. The two pulses are overlapped in space by assuring that the signal THz pulse strikes the polarizer at the same location as the idler THz pulse travels through the polarizer. The $CdWO_4$ sample is oriented at 0°, with the optical axis vertical.

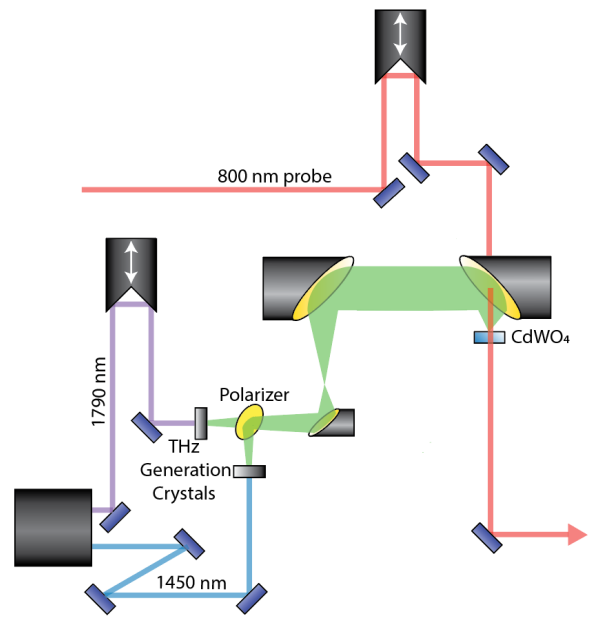

**Fig. S3. Experimental setup to perform 2D THz measurements with perpendicular pump polarizations**. Experimental setup showing the signal (blue, 1450 nm) and idler (purple, 1790 nm) OPA outputs traveling through DAST crystals to generate THz. The vertically polarized signal THz pulses reflect off a polarizer while the horizontally polarized idler travels through the polarizer, spatially overlapped with the signal. The two THz pulses travel colinearly through the parabolic mirror scheme to focus to the same spot on the $CdWO_4$ sample. The 800 nm probe pulse (red) is focused to the same spot on the sample and measured through heterodyne detection.

Fig. S4 displays the data from a 2D measurement where the THz pulses were perpendicularly polarized. Notice that the combined idler and signal sample response in Fig. S4A looks nearly identical to Fig. S4B. When the two THz pulses are orthogonally polarized, ERS excitation can only occur when the two THz electric fields are overlapped in time because of the symmetry requirement for two perpendicularly polarized photons. Anharmonic coupling can occur over a longer time period. The two orthogonal THz pulses each align with a different IR-active phonon mode ($B_u$ or $A_u$). By the symmetry selection rules, the Raman-active modes can only be excited via anharmonic coupling when both $B_u$ and $A_u$ IR-active modes are excited, which is also strongest at THz pump delays where the two pulses overlap. However, phonons do not damp out as soon as the THz pulses disappear so both $B_u$ and $A_u$ IR-active modes will remain excited for a time after the THz pulses cease to be overlapped, allowing for anharmonic coupling over a longer time range than the ERS. The nonlinear signal nearly matches the combined pump pulse signal because only nonlinear signal is present; the small differences in Fig. S4A and B arise from the pump polarizations not being perfectly vertical and horizontal.

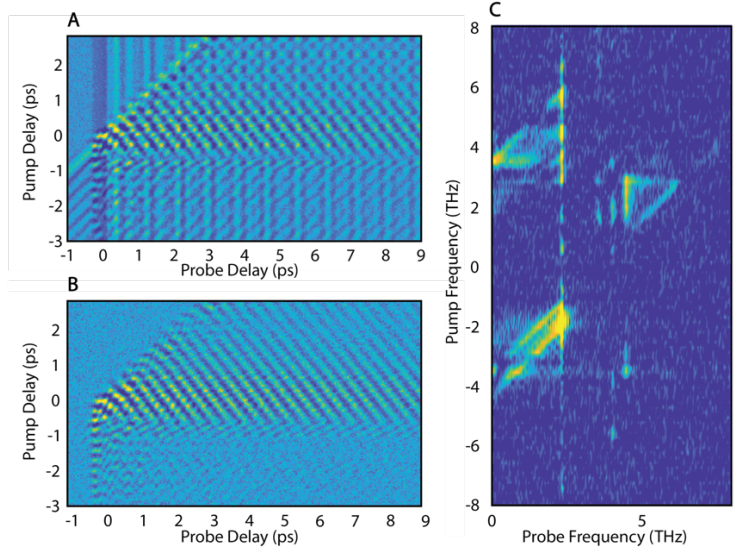

**Fig. S4. Time and frequency signals from 2D THz spectroscopy with perpendicular pump polarizations.** (**A**) A 2D plot of the time domain when both THz pulses are present. (**B**) The nonlinear time-domain signal isolated after chopping and (**C**) the resulting 2D frequency correlation spectrum.

The nonlinear 2D frequency spectrum in Fig. S4C has distinct differences from Fig. S2C. There are two major implications of the perpendicular 2D THz measurement. First, an ERS response can only occur when the two electric fields are overlapped in time. Secondly, both IR-active phonons are strongly driven to large amplitudes because the electric field polarization orientations are fully optimized to drive each IR mode involved in the trilinear coupling. The large motion in the IR-active modes produces a stronger trilinear coupling signal response from the coupled Raman-active mode.

S4 THz Field-Strength Dependence

We measured the field dependence of the 2D frequency correlation spectrum by changing the angle between two crossed polarizers in the parallel THz configuration described above. The 2D spectra at five different field strengths are pictured in Fig. S5. At the highest field strength, there are many clear features in the spectra. The amplitude of these features decreases as the field strength decreases. In Fig. S6, we show an example of one of these regions indicated by the red box in Fig. S6A. As shown by the red dots in Fig. S6B which are averages of the spectral amplitude in the area indicated in Fig. S6A, this feature follows a quadratic dependence; all other regions of the 2D spectrum with a significant spectral amplitude show the same quadratic dependence. This is an indicator that we do not need to include fourth order (or higher order) terms in our equations of motion to model the $CdWO_4$ response since fourth order equations would have a cubic (or higher order) dependence on the electric field ($E(t)^3$ or $\Phi_{abcd} Q_a Q_b Q_c Q_{R_d} \propto E(t)^3$). An example of this quadratic field dependence is shown in Fig. S6, where each red dot represents the average peak amplitude of the red selected area as a function of THz power.

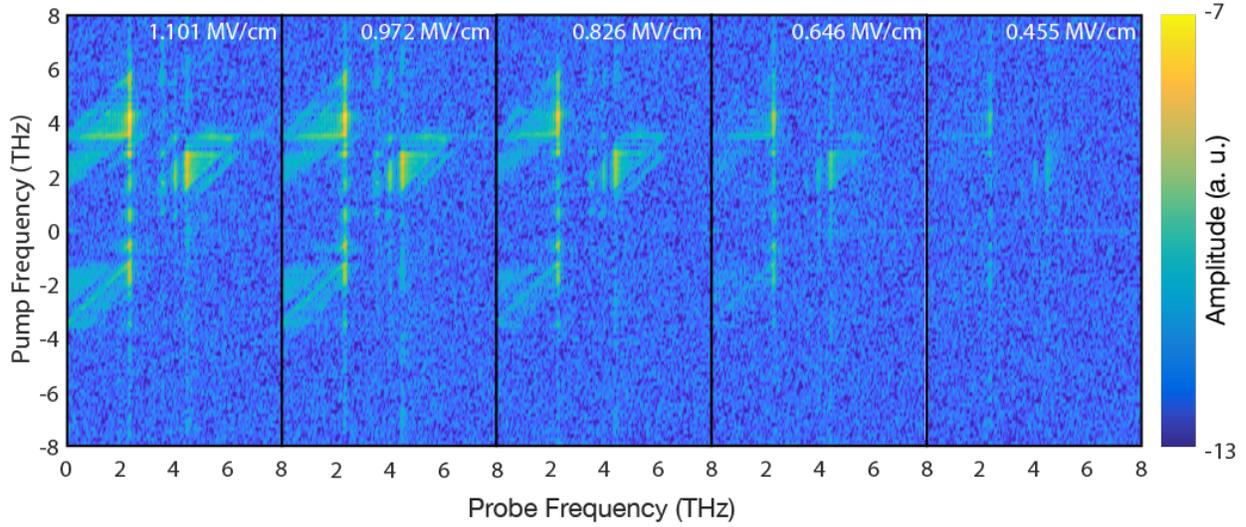

**Fig. S5. Two-Dimensional Field Dependence.** 2D frequency correlation spectra measured with THz field strengths of 1.101 MV/cm, 0.972 MV/cm, 0.826 MV/cm, 0.646 MV/cm, and 0.455 MV/cm. The yellow nonlinear signal decreases with field strength.

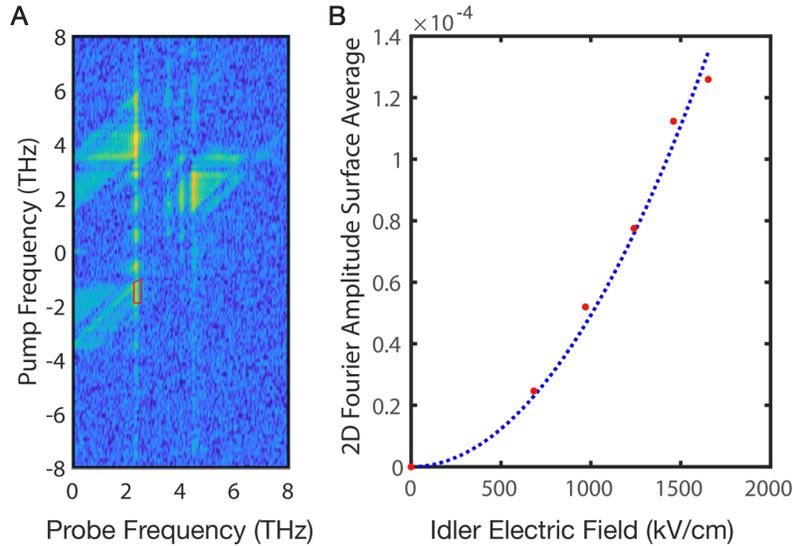

**Fig. S6. Quadradic field dependence of nonlinear signals.** (**A**) 2D frequency correlation spectrum with 1.101 MV/cm field strength. The red outline indicates one of the regions used to compare average amplitude at each power. (**B**) Field dependence of the averaged region outlined in red in **A**.

S5 Fitting

   We performed a manual Chi-squared fit to extract anharmonic coupling and ERS parameters from our 2D-THz measurements. We fitted four Raman scattering tensors ($R_{i\alpha\beta}$) and eight anharmonic coupling constants ($\Phi_{i,6,8}$ and $\Phi_{i,8,10}$), where $i$=4, 7, 9, 11. We used $\Phi_{4,6,8}$ as a reference by assigning it a value of 1 throughout the fitting as seen in Table I of Fig. S7. We then varied $\Phi_{4,8,10}$ over a broad range of possible values and created a modeled spectrum for each value of $\Phi_{4,8,10}$. While we did this, all other parameters were set to zero. We calculated the RMSE of the Fourier amplitude of our experimental frequency correlation spectrum with our modeled spectrum for each value, while keeping the remaining parameters as zeroes (Fig. S7A). This allowed us to find an optimal value for $\Phi_{4,8,10}$ at the minimum RMSE of the Fourier amplitudes (in this case -2.2), which was kept in our model. We modeled the spectrum with only $\Phi_{4,8,10}$ and $\Phi_{4,6,8}$ as nonzero values in Fig. S7B. Through the same process, we determined a value for $\delta_{R_4}$, this time by varying the value of $R_{4\alpha\beta}$, and leaving only $\Phi_{4,8,10}$ and $\Phi_{4,6,8}$ as nonzero (see Table II of Fig. S7). The resulting RMSE shown in Fig. S7C minimized to a value of -10 for $R_{4\alpha\beta}$. In Fig. S7D, we then produced a model with $\Phi_{4,8,10}$, $\Phi_{4,6,8}$, and $R_{4\alpha\beta}$ as the nonzero parameters. Following this same method, we determined a value of -6 for $R_{7\alpha\beta}$ in Fig. S7E and modeled the spectrum (Fig. S7F) with this new value along with the other values in Table III of Fig. S7. The newly determined values were kept in the model as the new parameters, replacing the zero values.

This pattern of determining the anharmonic coupling constants and their related Raman scattering tensor through the RMSE of the Fourier amplitude was followed for the remaining parameters; however, in those cases their Raman scattering tensor was optimized first and then their anharmonic coupling constants. In the end, all the Raman scattering tensors and anharmonic coupling constants were fitted (see Table IV of Fig. S7) and the determined values were used as the new parameters to model a frequency correlation spectrum Fig. S7G and H.

We performed this fit two times. First, we minimized the RMSE by using the absolute value of the Fourier amplitude to simultaneously affect all the features in the frequency correlation spectrum (see Table IV of Fig. S7 for the values of the first fit). Second, we took the natural log of the absolute value of the Fourier amplitude in the RMSE in order to enhance the fitting of the smaller features present in the frequency correlation spectrum. In the second fitting, the starting values of all the parameters were the previously determined values from the first fit, rather than starting at values of zero. After the two fits, a set of optimized anharmonic coupling constants and Raman scattering tensors were determined and used to model a frequency correlation spectrum (Fig. S7I). This newly modeled frequency correlation spectrum was compared with our experimental plot (Fig. S7J). The modeled spectrum and the experimental spectrum have many features in common. The extracted Raman scattering tensors and the anharmonic coupling constants were numerically compared (Table S6) to those obtained from first principles calculations.

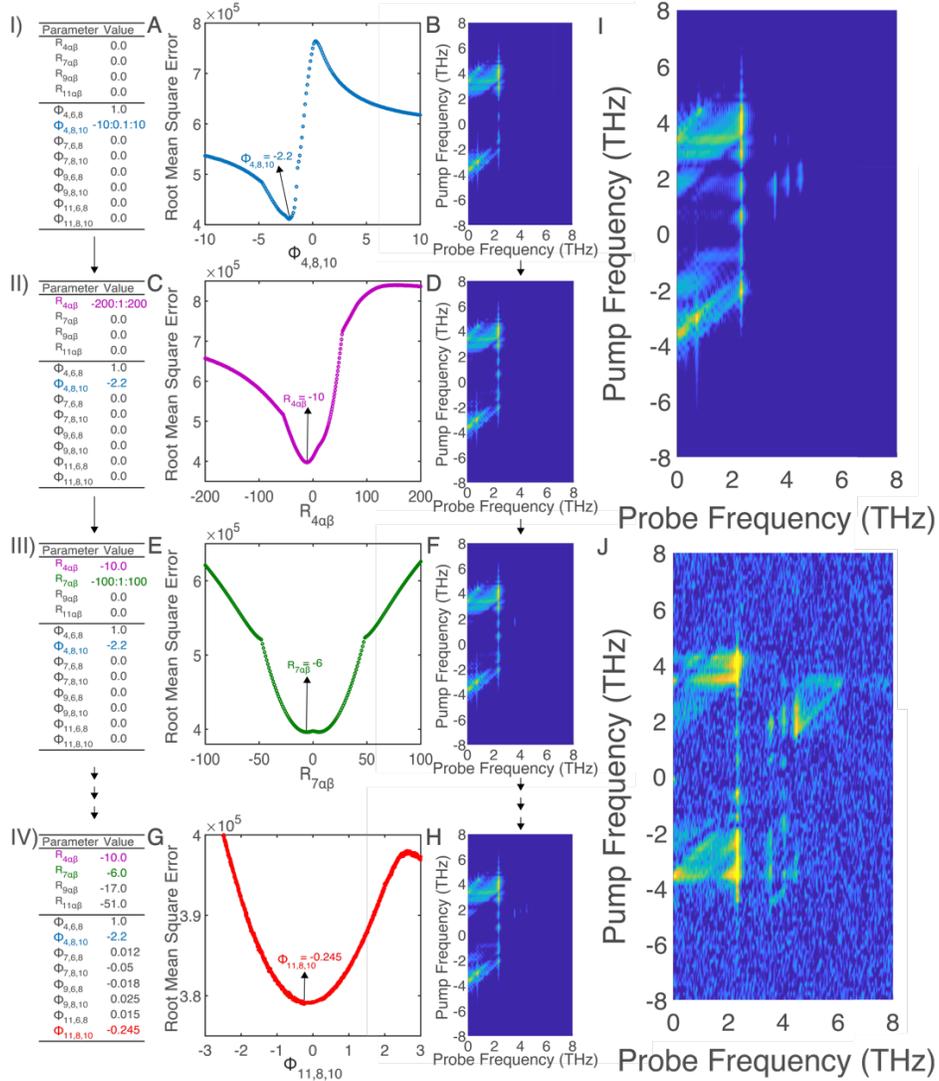

**Fig. S7. Step-by-step description of the fitting process.** (I) Values of the anharmonic coupling and ERS parameters used to minimize the RMSE of $\Phi_{4,8,10}$ along with the range of $\Phi_{4,8,10}$ values used for the fit. (**A**) RMSE used to extract the value of $\Phi_{4,8,10}$ in the first fitting. (**B**) Modeled frequency correlation spectrum for CdWO$_4$ using fitted value of $\Phi_{4,8,10}$. (II) Values of parameters used to minimize the RMSE of $R_{4\alpha\beta}$ along with its range of values. (**C**) RMSE used to extract the value of $R_{4\alpha\beta}$ in the first fitting. (**D**) Modeled frequency correlation spectrum for CdWO$_4$ using fitted value of $\Phi_{4,8,10}$ and $R_{4\alpha\beta}$. (III) Values of parameters used to minimize the RMSE of $R_{7\alpha\beta}$ along with its range of values. (**E**) RMSE used to extract the value of $R_{7\alpha\beta}$ in the first fitting. (**F**) Modeled frequency correlation spectrum for CdWO$_4$ using fitted value of $\Phi_{4,8,10}$, $R_{4\alpha\beta}$, and $R_{7\alpha\beta}$. (IV) Final determined set of values of the anharmonic coupling and ERS parameters after minimizing the RMSE of each parameter for the first fitting. (**G**) Final RMSE plot of the first fitting, extracting $\Phi_{11,8,10}$. (**H**) Modeled frequency correlation spectrum including all of the extracted $R_{i\alpha\beta}$ and $\Phi_{i,j,k}$ values from the first fitting. (**I**) Final modeled frequency correlation spectrum after the second fitting. (**J**) Experimental frequency correlation spectrum.

| Parameter | Value |
|---|---|
| $R_{4\alpha\beta}$ | -22.5 ± 12.5 |
| $R_{7\alpha\beta}$ | -2.8 ± 9.9 |
| $R_{9\alpha\beta}$ | 4.5 ± 65.3 |
| $R_{11\alpha\beta}$ | -39.0 ± 142.9 |
| $\phi_{4,6,8}$ | 1.00 |
| $\phi_{4,8,10}$ | -1.2 ± 0.6 |
| $\phi_{7,6,8}$ | 0.4 ± 0.2 |
| $\phi_{7,8,10}$ | -0.8 ± 0.5 |
| $\phi_{9,6,8}$ | 1.9 ± 1.2 |
| $\phi_{9,8,10}$ | -4.1 ± 3.4 |
| $\phi_{11,6,8}$ | 1.7 ± 1.4 |
| $\phi_{11,8,10}$ | -3.0 ± 1.5 |

**Table S6. Values of Raman tensor and trilinear coupling constants determined through fit.**

S6 Shutting off Electronic Raman Scattering

Modifications to the setup in Ref. (*8*) allow us to isolate trilinear phonon coupling that we were unable to see previously. As described above in Methods and Materials and section S3, the 2D THz setup incorporates strong variably delayed THz pulses, chopping to isolate nonlinear signal, and polarization control of the individual THz pump beams. Modeling shows that crossed polarization of the THz pump pulses eliminates the presence of ERS signal when the THz pump pulses are not overlapped in time. Fig. S8 shows a side-by-side comparison of the modeled excitation of Raman mode 4 when driven by ERS or trilinear coupling in the scenario of either parallel or perpendicular THz pulse polarizations. As shown in the top 2 panels, when both pump pulses have the same polarization and the sample is oriented at 45 degrees, we see ERS and anharmonic coupling excitation when the pump pulses are not overlapped in time. However, when the pump pulses have perpendicular polarization and the sample is oriented at 0 degree azimuthal angle, the Raman mode is not excited by ERS when the pump pulses are not overlapped in time (see bottom left panel of Fig. S8). However, the bottom right panel shows that anharmonic coupling excitation is increased compared to the parallel excitation. This modeling shows that perpendicularly polarized THz pulses allow us to more cleanly isolate trilinear coupling signal that we were not able to see in Ref. (*8*).

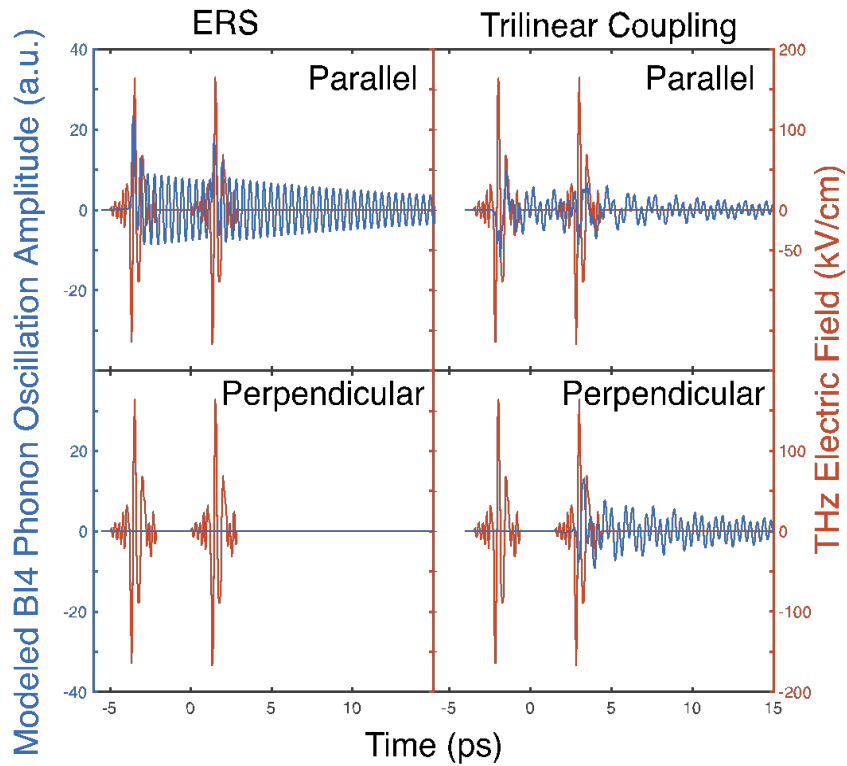

**Fig. S8. Nonlinear ERS and trilinear coupling signals depend on pump polarization.** Modeled oscillations for the Raman-active mode 4 (blue) are compared after being driven by two non-overlapped THz (red) pump pulses. The THz pulse polarizations are both parallel in the top panels and perpendicular in the bottom panels. Mode 4 is driven solely by ERS in left plots, and trilinear coupling in the right plots. ERS is essentially turned off when the THz pulses are not overlapped and have perpendicular polarizations, whereas anharmonic coupling excitation is maximized.

In measurements with parallel THz pulse polarizations with the sample at 45 degrees as in Ref. (*8*), Raman-active modes could be excited via ERS at all THz pump pulse delays because a single THz pulse contained both of the electric field polarization vector components required by the symmetry for ERS to occur. A single THz pulse provides both photons needed for an ERS excitation of the Raman-active modes. By changing the polarization of the two THz pulses to be orthogonal, the only way we can meet the symmetry requirements for ERS is if two photons with orthogonal light polarizations from the two THz pulses are spatially and temporally overlapped. This dramatically reduces the amount of signal from ERS. Although ERS still happens and dominates at delays where the THz pulses are overlapped, it allows us to observe previously masked nonlinear signal arising when the two pulses are not directly overlapped. We essentially turn off ERS with cross polarization at some delays, which means the ERS excitation mechanism contributes less to our nonlinear signal and no longer masks nonlinear signal from the trilinear phonon coupling mechanism.

Similar to ERS excitation, for trilinear phonon coupling to occur, both *phonons* must be excited at the same time. Since the phonon oscillations last many picoseconds (longer than the THz pulse envelope of photons), we can excite Raman-active modes via trilinear coupling at all experimental pump delays and observe the resulting nonlinear phononic signal.

*Electronic Raman Scattering in 2D correlation spectra*

We model the ERS signal for only mode 4 by setting all of the coupling constants and the other Raman tensor elements to zero leaving just Eq. S3.

$$\ddot{Q}_{R_4} + 2\Gamma_{R_4}\dot{Q}_{R_4} + \omega_{R_4}^2 Q_{R_4} = -\varepsilon_0 R_{4\alpha\beta} E_\alpha(t) E_\beta(t) \qquad \text{Eq. S3}$$

Fig. S9A shows our model for just the Raman excitation of mode 4 using Eq. S3. The Raman excitation is a vertical stripe in the data at the resonant frequency of the Raman-active mode being excited, in this case at 2.33 THz. The pump axis gives information about what frequency elements were involved in producing the nonlinear excitation. The probe axis indicates which modes, visible to the probe pulse, were excited nonlinearly. The vertical stripe at 2.33 THz on the probe axis indicates that many different pump-pulse frequency combinations were involved in exciting the Raman-active mode 4. In our data we also see a vertical stripe at the frequency of mode 4 (Fig. S9B) as well as at the frequencies of the other three Raman-active $B_g$ modes in $CdWO_4$, indicating that all four $B_g$ Raman-active modes are being excited nonlinearly through ERS.

Fig. S10A shows a normalized DAST THz generation crystal spectrum. For THz-ERS excitation to occur, two photons must combine whose frequencies either add up to or subtract down to the frequency of the Raman-active mode they are exciting. Vertical lines and associated colored circles in Fig. S10 show four possible combinations of photons included in our pump pulse spectrum whose frequencies add up or subtract down correctly to excite mode 4 in $CdWO_4$. Fig. S10B indicates where several of these photonic pairs would appear on the nonlinear 2D spectrum for $CdWO_4$, where the frequencies of the photons being combined appear along the pump axis and at $v_{R4}$ along the probe axis. With these four examples, we can see a vertical line beginning to form and one can imagine that including all the possible photonic combinations would completely fill in the line. There are brighter and darker areas in the vertical stripe. These are caused by the different intensities of THz light in our pump at different frequencies. We also see clear brighter features in the experimental spectrum that arise from anharmonic coupling.

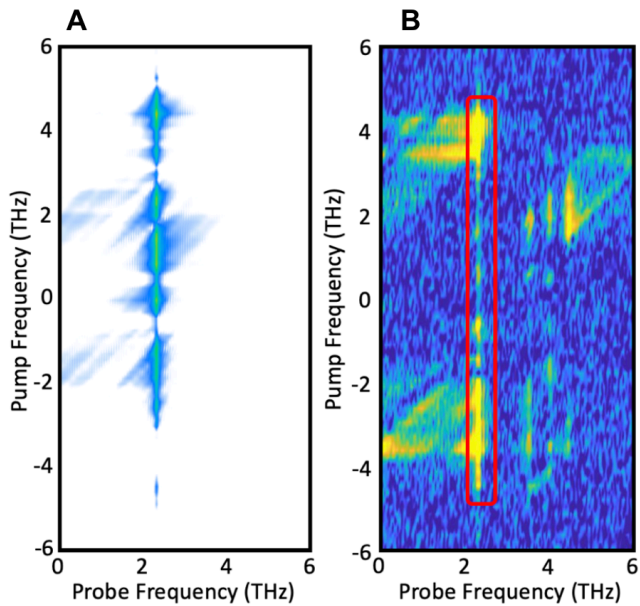

**Fig. S9. Modeled and experimental ERS signals.** (**A**) Modeled 2D correlation spectrum for the ERS excitation of mode 4. (**B**) Experimental frequency correlation spectrum showing the vertical ERS stripe boxed in red.

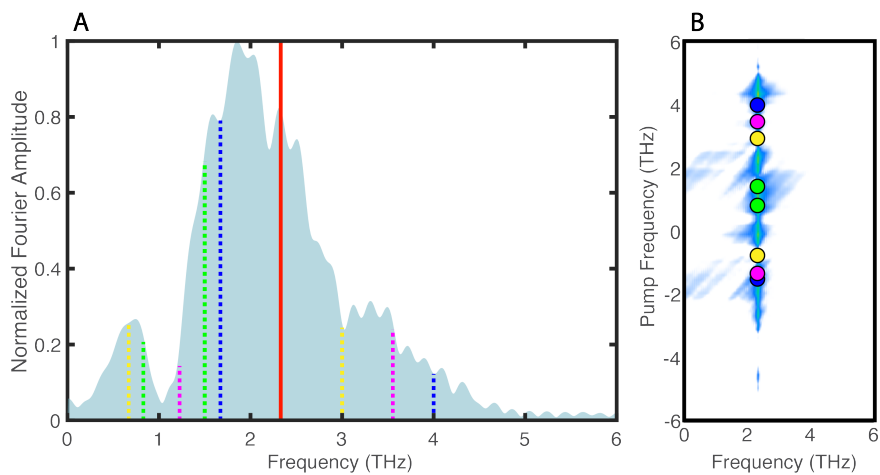

**Fig. S10. ERS is a vertical stripe in the data because of various possible frequency combinations.** (**A**) Pump-pulse THz spectrum (light grey). The red line indicates the frequency of the Raman-active mode being excited through ERS. The dotted lines indicate pairs of frequencies included in the pump pulse that can add up to or subtract down to the frequency of the Raman-active mode. (**B**) Modeled frequency correlation spectrum for the ERS excitation of mode 4, showing the signal as a vertical stripe. The colored dots indicate where the signals created by the ERS frequencies indicated in **A** appear in the frequency correlation spectrum, explaining in part why the ERS signal shows up as a vertical stripe.

S7 Modeled Anharmonic Coupling Peak Assignment

Using modeling with the coupled equations of motion for arbitrary modes, we were able to identify a framework where anharmonic coupling features appear in a typical 2D frequency correlation spectrum. Along the probe frequency axis, anharmonic coupling features appear at the frequency of the Raman-active mode involved in the coupling, which is the mode that we can probe. Along the pump frequency axis, we see features at the resonant frequencies of the IR-active modes involved in the coupling as well as at sum and difference frequencies between the individual IR-active modes and the Raman-active modes (Fig. S11). As an example, in Fig. S11A we have a modeled spectrum with arbitrary phonon frequencies at 0.9 THz, 1.3 THz, and 2.7 THz. We are probing the 1.3 THz Raman-active mode, so all features will appear on that line on the probe axis. Along the pump frequency axis, we marked with squares the features appearing at the frequency of the two IR-active modes involved in the coupling: 0.9 THz and 2.7 THz. Difference frequency features, shown in circles, appear at 1.3 - 0.9 = 0.4 THz and 1.3 - 2.7 = -1.4 THz on the pump axis. Sum frequency features, shown in triangles, appear at 1.3 + 0.9 = 2.2 THz and 1.3 + 2.7 = 4 THz on the pump axis.

In Fig. S11B and C, we have modeled the three phonon modes in $CdWO_4$ at 2.3 THz, 2.9 THz and 3.6 THz. The 2.3 THz mode is a Raman-active $B_g$ mode, and this is the mode that is probed with our setup. The 2.9 and 3.6 THz modes are IR-active $B_u$ and $A_u$ modes, respectively. Difference frequency features appear at -1.3 THz and -0.6 THz, and are shown in circles. Sum frequency features appear at 5.2 THz and 5.9 THz and are again shown in triangles. Fig. S11B is modeled with the polarization of the "stationary" THz pulse vertically polarized and the "variable delay" THz pulse horizontally polarized, while Fig. S11C has the stationary pulse horizontally polarized and the variable delay pulse vertically polarized. Modeling using these different angles highlights different features because of the symmetry of the modes that each pulse excites. If one of the features is less prominent or does not appear, the shapes that highlight these features are dashed. We note, however, that if a feature does not appear in one figure, it usually appears in the other.

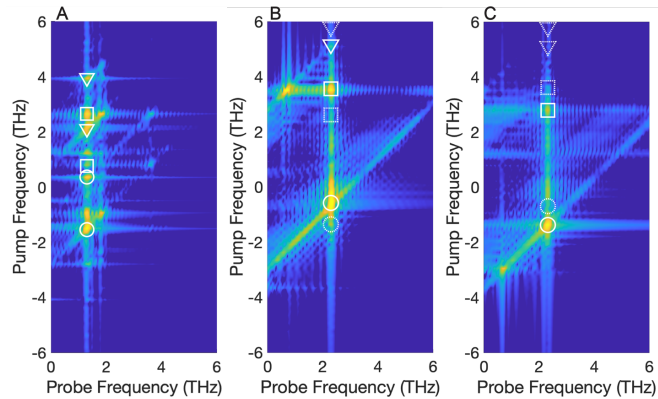

**Fig. S11. Anharmonic coupling signals show up at specific resonant, sum, and difference frequencies.** Modeled frequency correlation spectra of (**A**) arbitrary frequencies and (**B**), (**C**) $CdWO_4$ frequencies showing features at mode frequencies (squares), difference frequencies (circles) and sum frequencies (triangles). (**B**) is modeled with the "stationary" THz pulse having vertical polarization and the "variable delay" pulse with horizontal polarization, while (**C**) is modeled with a horizontally polarized stationary pulse and a vertically polarized variable delay pulse.

S8 First Principles Calculations

First-principles calculations were performed to calculate the anharmonic coupling constants and Raman tensor elements for the relevant modes in CdWO$_4$. The first-principles parameters for CdWO$_4$ were obtained using the Vienna ab initio simulation package (VASP). Throughout this paragraph, the INCAR tags used in VASP are listed in all caps in parentheses. This involved first relaxing the crystal structure with a conjugate gradient algorithm (IBRION = 2) until the norms of all the Hellmann-Feynman forces were smaller than 0.0001 eV (-EDIFFG) and the ions were in their instantaneous ground state. The cutoff energy for the plane-wave-basis was set to 1000 eV (ENCUT). We used an 8x8x8 k-point Monkhorst-Pack mesh centered at the gamma point to sample the Brillouin zone. We used the solid Perdew-Burke-Ernzerhof (PBEsol) form of the generalized gradient approximation (GGA) for the exchange-correlation functional. Once the relaxed structure was obtained, we employed the finite difference method (IBRION = 6) to calculate the forces of the relaxed and displaced unit cell structures (pictured below, Fig. 12(**A**)).

Our method requires only 2n+1 calculations in order to find all possible mode couplings of the 3rd and 4th order, where n is the number of modes in the material, in contrast to the thousands of calculations required to find a single mode coupling using the finite difference method. First, we perform a finite difference calculation on the relaxed structure, and then we perform finite difference calculations on one positively and one negatively displaced structure along each mode coordinate, for a total of 2n+1 calculations. These calculations output the Hessian matrices, eigenvectors and eigenfrequencies (Fig. 12B and C), Born effective charges, and static dielectric constants of CdWO$_4$. In Table S7 we list the calculated frequencies and experimental damping rates of our modes of interest. The anharmonic force constants can be extracted from the Hessian matrix ($\Phi_{ij}$) using the Equations S4-6. These $\Phi$s are in mode coordinates rather than atomic coordinates.

$$\Phi_{ijk} = \frac{1}{3}\left(\frac{\Phi_{jk}(\delta Q_i) - \Phi_{jk}(-\delta Q_i)}{2\delta Q_i} + \frac{\Phi_{ki}(\delta Q_j) - \Phi_{ki}(-\delta Q_j)}{2\delta Q_j} + \frac{\Phi_{ij}(\delta Q_k) - \Phi_{ij}(-\delta Q_k)}{2\delta Q_k}\right)$$

Eq. S4

$$\Phi_{ijkk} = \frac{\Phi_{ij}(\delta Q_k) + \Phi_{ij}(-\delta Q_k) - 2\Phi_{ij}(0)}{\delta Q_k^2}$$

Eq. S5

$$\Phi_{iikk} = \frac{1}{2}\left(\frac{\Phi_{ii}(\delta Q_k) + \Phi_{ii}(-\delta Q_k) - 2\Phi_{ii}(0)}{\delta Q_k^2} + \frac{\Phi_{kk}(\delta Q_i) + \Phi_{kk}(-\delta Q_i) - 2\Phi_{kk}(0)}{\delta Q_i^2}\right)$$

Eq. S6

Some specific experimental details needed to be considered to calculate the coupling constants and Raman tensor elements of CdWO$_4$ from first-principles calculations. First, the THz pump driving force undergoes many corrections. We account for THz reflective losses off the surface of CdWO$_4$ by calculating the dielectric constant from first-principles. We then use the dielectric constant to calculate the field transmission coefficient in the material, specific to our crystal orientation. Thinking about the THz driving force past the sample surface, we need to calculate the IR mode frequencies and mode effective charge of each IR mode. This allows us to input a more accurate THz driving force into our model. Next, we use a finite difference approach to calculate the coupling constants which govern the flow of energy between phonon modes. We solve for the Raman Tensor at THz frequencies which allows us to account for nonlinear excitation of Raman active modes through a two-photon process. We also calculate the Raman Tensor at the probe frequency (800 nm) to know how sensitive the probe is to each Raman active phonon. Additionally, we calculate the mode effective charge and frequencies from finite difference calculation from a relaxed structure. The Raman tensors and coupling constants come from numerical derivatives on forces from displaced structures and the damping rates are added in manually.

| Band Index | Calculated (THz) | Experimental (THz) | Type | Axis | Exp. Damping Rate (THz) |
|---|---|---|---|---|---|
| 1-3 | - | - | Acoustic | - | - |
| 4 | 2.27 | 2.33 | R ($B_g$) | - | 0.061 |
| 5 | 3.01 | 2.90 | R ($A_g$) | - | - |
| 6 | 3.07 | 2.94 | IR ($B_u$) | x, y | 0.1049 |
| 7 | 3.45 | 3.56 | R ($B_g$) | - | 0.05 |
| 8 | 3.83 | 3.65 | IR ($A_u$) | z | 0.06 |
| 9 | 3.95 | 4.04 | R ($B_g$) | - | 0.207 |
| 10 | 4.35 | 4.47 | IR ($B_u$) | x, y | 0.27 |
| 11 | 4.44 | 4.48 | R ($B_g$) | - | 0.1709 |

**Table S7. First-principles Calculated and Literature Parameters. Damping rates from Refs (*8,14*).**

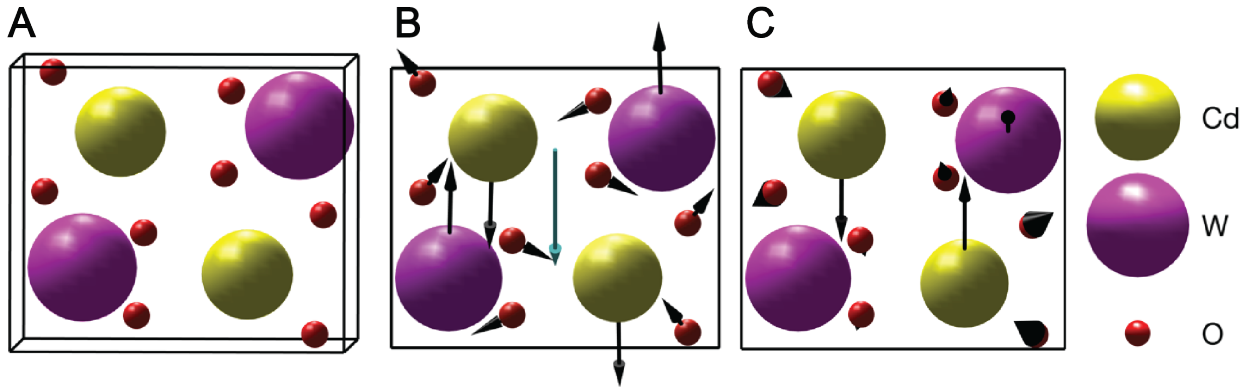

**Fig. S12. CdWO$_4$ unit cells with different mode vectors.** (**A**) Relaxed atomic positions of CdWO$_4$. (**B**) Unit cell of CdWO$_4$ including the calculated eigenvectors of an excited IR-active vibrational mode, with the blue arrow showing the mode effective charge vector. (**C**) Unit cell of CdWO$_4$ including the calculated eigenvectors of an excited Raman-active vibrational mode.